\documentclass[runningheads]{svmult}

\usepackage{makeidx}   
\usepackage{graphicx}  
\usepackage{subeqnar}  
\usepackage{multicol}  
\usepackage{physprbb}  
\makeindex             

\newcommand{\arcsec}{\hbox{$^{\prime\prime}$}}
\newcommand{\micron}{\,\hbox{$\mu$m}}
\newcommand{\lsun}{\,\hbox{L$_{\odot}$}}
\newcommand{\msun}{\,\hbox{M$_{\odot}$}}
\newcommand{\kms}{\,\hbox{km\,s$^{-1}$}}

\begin{document}
\title*{Nuclear Dynamics and Star Formation of AGN}
\toctitle{Nuclear Dynamics and Star Formation of AGN}
%
%
\titlerunning{Nuclear Dynamics and Star Formation of AGN}
%
\author{Richard Davies\inst{1}
\and Linda Tacconi\inst{1}
\and Reinhard Genzel\inst{1}
\and Niranjan Thatte\inst{2}}
\authorrunning{Davies et al.}
%
%
\institute{Max-Planck-Institut f\"ur extraterrestrische Physik,
Postfach 1312, 85741 Garching, Germany
\and Department of Astrophysics, Denys Wilkinson Building, Keble Road,
Oxford, OX1 3RH, UK}

\maketitle              

\begin{abstract}
We are using adaptive optics on Keck
and the VLT to probe the dynamics and star formation in Seyfert and
QSO nuclei, obtaining spatial resolutions better than 0.1\arcsec\ in
the H- and K-bands.
The dynamics are traced via the 2.12\micron\ H$_2$ 1-0\,S(1) line, while
the stellar cluster is traced through the CO\,2-0 and 6-3 absorption
bandheads at 2.29\micron\ and 1.62\micron\ respectively.
Matching disk models to the H$_2$ rotation curves allows us to study
nuclear rings, bars, and warps; and to constrain the mass 
of the central black hole.
The spatial extent and equivalent width of the stellar absorption
permits us to estimate the mass of stars in the nucleus and their
contribution to the emission.
Here we report on new data for I\,Zwicky\,1, Markarian\,231, and
NGC\,7469.
\end{abstract}

\section{Introduction}

QSOs include some of the most luminous objects in the universe.
Their prodigious energy output is powered by accretion onto a black
hole with mass in the range $10^6$--$10^9$\msun, although it is
becoming increasingly apparent that star formation also plays an
important part --- one that is perhaps still underestimated.
This project focusses on studying the gas and stars in the nuclei of
such AGN. 
In order to select nearby targets where adaptive optics can probe the
nuclear scales, we are limited to the lower end of the luminosity
range spanning the crossover between QSOs and Seyfert nuclei.
The aims are to: 
1) measure the gas dynamics on scales less than 1\,kpc
to understand how gas is driven in to the nucleus, and to
constrain the mass of the black hole there;
2) determine the contribution and mass of the nuclear stellar
cluster, and to further our understanding of the relation
between an AGN and the surrounding star formation.

In this summary, we briefly discuss 3 nearby QSO/Seyfert nuclei for
which at least some data has been analysed:
spectroscopy of NGC\,7469 and Mkn\,231, and imaging of I\,Zw\,1.
Even for these objects, 1\arcsec\ corresponds to 300-1200\,pc, making the
use of adaptive optics mandatory to probe the nuclear scales of order
100\,pc or less.
These are all luminous objects, with $L_{\rm IR} =
3\times10^{11}$--$3\times10^{12}$\lsun, in which the nuclear activity 
appears to have been triggered by an interaction.
In every case at least 2/3 of the luminosity
originates in star formation rather than the black hole.
So it is clear that star formation does play a crucial role in AGN, and
we must begin with such objects if we are to understand 
how this may apply to the more luminous QSOs at higher redshift.

\begin{table}
\caption{Observations to Date}
\begin{center}
\renewcommand{\arraystretch}{1.4}
\setlength\tabcolsep{5pt}
\begin{tabular}{llllll}
\hline\noalign{\smallskip}
Object & Telescope & Instrument & Slit Width & Band & Resolution \\
\noalign{\smallskip}
\hline
\noalign{\smallskip}
NGC 5506 & VLT  & NAOS + CONICA & 86\,mas & K & 1400 \\
I Zw 1   & VLT  & NAOS + CONICA & 86\,mas & H/K & 1500/1400 \\
Mkn 231  & Keck & NIRC-2 + AO   & 80\,mas & H/K & 1800/2500 \\
Mkn 509  & Keck & NIRC-2 + AO   & 80\,mas & K & 2500 \\
NGC 7469 & Keck & NIRSPEC + AO  & 37\,mas & K & 2900 \\
\hline
\end{tabular}
\end{center}
\label{tab:obs}
\end{table}

\section{Instrumentation and Observations}

A summary of the spectroscopic observations which have been made to
date (from mid-2002 until late-2003) is given in Table~\ref{tab:obs}.
The data have been collected using both the Keck\,II
telescope and the VLT, putting
us in a fortunate position of being able to compare the
adaptive optics systems and their instrumentation.
The original AO camera on Keck was NIRSPEC which used a
special optical feed to change the plate scale so that it was
suitable for adaptive optics.
Although this meant that the slit length was only 4\arcsec\ and the
width 37--74\,mas, it was possible to achieve spectral resolutions
of $R\sim2500$ necessary for dynamics work.
This is now superceeded by NIRC-2 which is designed to work with an AO
system.
Its larger detector provides good sampling
with a 10--40\arcsec\ field of view, and allows one to reach the
necessary spectral resolution with a wider slit, better matched to the
full size of the diffraction limited PSF.
While the AO system is good and relatively straight forward to use,
it has only a fixed number of about $18\times18$ subapertures.
This is where NAOS on the VLT has a big advantage: it can switch
between $14\times14$ and $7\times7$ lenslet arrays, allowing it to
perform well even on fainter objects. 
For example, in unexceptional conditions it was possible to reach 15\%
Strehl in the H-band correcting on the slightly extended $V\sim14.1$
nucleus of I\,Zw\,1.
While its camera CONICA has a large number of observational modes, and
is equipped with tools that make it easy to align
the slit across 2 objects, the spectral resolution is relatively low.
It is not yet clear whether $R\sim1500$ (200\kms) is sufficient
to allow useful dynamics studies of AGN.

\section{I Zw 1}

\begin{figure}
\centerline{\includegraphics[width=0.9\textwidth]{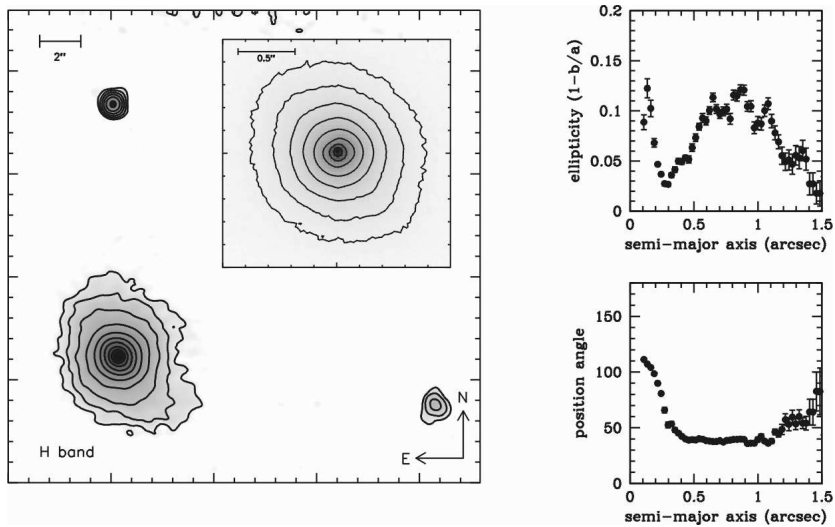}}
\caption{I\,Zw\,1. Left: H-band image of a 23\arcsec\ FoV showing 2
off-nuclear emission regions (adaptively smoothed; contours at factor
2 intervals); inset is an expanded view of the central 2\arcsec.
Right: ellipticity and position angle fits to
the central isophotes.}
\label{fig:izw1}
\end{figure}

An H-band image of I\,Zw\,1 is presented in Fig.~\ref{fig:izw1}.
In addition to the bright nucleus, two outlying regions are visible.
Both of these, as well as fainter emission throughout the whole
region and a tidal tail have been seen at other wavelengths
(e.g. J-band, see \cite{sch03}).
The faint diffuse region to the west has usually been
interpreted as a tidal dwarf galaxy;
the compact northern knot was thought to be a foreground star.
We have now resolved the northern region (transverse to its offset
from the nucleus, to minimise anisoplanatic effects) to have a FWHM of
0.186\arcsec, equivalent to 220\,pc.
We have not yet measured the flux density, nor estimated
its stellar mass --- but the large size already suggests that it may
also be a tidal dwarf galaxy, or possibly even the progenitor nucleus
of the galaxy with which I\,Zw\,1 is interacting.

The inset in Fig.~\ref{fig:izw1} reveals elliptical isophotes
around the AGN.
These are quantified in the right hand panels: at
radii 0.6--1.1\arcsec\ the axis ratio is 0.9 at a position angle (PA)
of $40^\circ$.
This is perhaps surprising, since radio CO\,1-0 data in the central
few arcsec \cite{sch98} show that the
nucleus is inclined by $38\pm5^\circ$ at PA $135^\circ$ ---
almost perpendicular to the elongation of the isophotes.
Deprojecting the nuclear region thus increases the isophotal axis
ratio to about 1.4.
Such a strong deviation from axisymmetry implies the presence of a
stellar bar on scales of 0.5--1\,kpc, which may be the key to
understanding how gas is fed into the nuclear region to fuel the AGN
and star formation there.

Lastly, the nucleus itself has a FWHM of 0.102\arcsec, 
although the data are diffraction limited (the Strehl ratio
achieved was 15\%).
Even accounting for loss of resolution due to undersampling
and sub-pixel shifting, we have resolved the nucleus.
This suggests that there is 
a significant stellar cluster on scales of 100\,pc.

\section{Mkn 231}

\begin{figure}
\centerline{\includegraphics[width=1.0\textwidth]{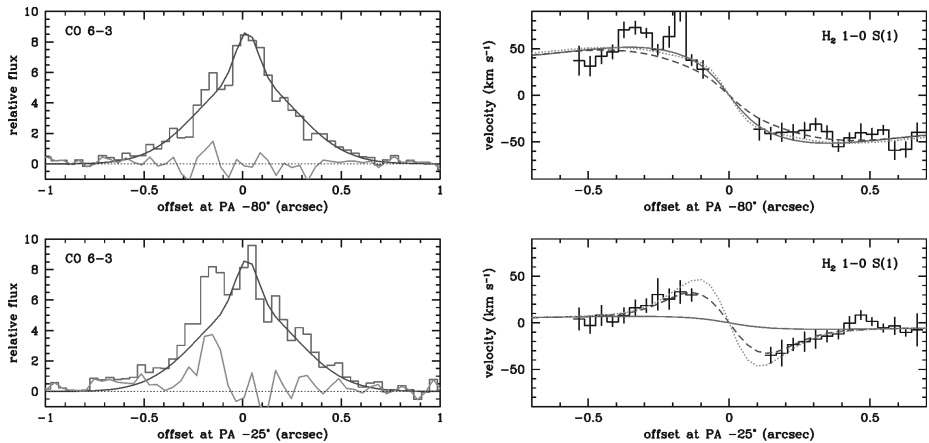}}
\caption{Mkn\,231. Left: spatial profiles of the 
1.62\micron\ CO\,6-3 stellar absorption (stepped line)
at 2 position angles.
Both plots show the same 2-Gaussian fit to the CO profile
(dark curve), and the associated residuals (light curve).
Right: velocity profiles of the 2.12\micron\ H$_2$ 1-0\,S(1) line
(solid stepped line, 1$\sigma$ errors), with 3 models
overplotted.}
\label{fig:mkn231}
\end{figure}

Fig.~\ref{fig:mkn231} summarises the spectroscopic results for
Mkn\,231 (see \cite{dav04b} for a more complete analysis).
The left hand panels show the spatial profile of the 1.62\micron\
CO\,6-3 stellar absorption, which has been fit with a double
Gaussian.
At $-80^\circ$, the stellar absorption is extended with a FWHM of
0.37\arcsec.
For reference, the H-band continuum has a FWHM of 0.19\arcsec, and the
resolution is smaller still.
The stellar cluster is therefore resolved over a scale of
300\,pc.
At $-25^\circ$ the CO profile can be reproduced by the same double
Gaussian, apart from a residual bump (which also appears in the
continuum) offset by $-0.15$\arcsec\ (120\,pc) from the peak
This cannot be an AO artifact because its shape is
well fit in both continuum and CO by a Gaussian with FWHM 0.10\arcsec,
rather than simply being a copy of the on-axis shapes.
Also, the equivalent width of the CO feature ($W_{CO}$) is different.
In the core of the on-axis peak $W_{CO}=0.15$\,\AA\ (compared to late
type stars which have $W_{CO}=3$--5\,\AA), while
the off-axis feature has $W_{CO}=0.69$\,\AA;
the extended CO has a mean value $W_{CO}=0.76$\,\AA.
A full analysis of the mass of the nuclear stellar cluster has not
yet been been carried out, but based on these numbers it is already
clear that the CO feature is hugely diluted by 
hot dust associated with the AGN (and perhaps also to some extent with
the star cluster itself):
even in the H-band the on-axis dilution is a factor 20.

The velocity profiles of the 2.12\micron\ H$_2$ 1-0\,S(1) are plotted
in the two right hand panels, overdrawn with 3 models based on a thin
axisymmetric disk.
Model~1 (solid line) is for a single disk having a Gaussian
mass surface density with FWHM 0.5\arcsec.
The disk is inclined by $20^\circ$ with major axis at PA $-108^\circ$,
as determined by 0.6\arcsec\ radio CO\,2-1 measurements \cite{dow98}.
The mass has been set at $2.5\times10^9$\msun\ in order to match
the rotation curve at $-80^\circ$.
This model clearly does not replicate the Keplerian-like rotation
curve at $-25^\circ$.
Instead the observations indicate that at smaller radii gas exists
with a different effective major axis, requiring a warped disk model.
Model~2 (dashed line) reproduces this in a simplified way. 
At radii larger than $\sim0.15$\arcsec, the disk is as for Model~1.
But inside this, the disk is tilted out of its primary plane
until the observed major axis lies along $-25^\circ$.
In the model shown here, the resulting inclination of this inner disk
is $27^\circ$, although any value in the range 20--$40^\circ$
is possible depending on how much and at what PA the inner
disk is tilted.
This model matches the observations well, and tends to confirm that
there is indeed a warp in the central 100--150\,pc of Mkn\,231.
Model~3 (dotted line) shows the effect on the rotation curve if
there were also a $2.5\times10^8$\msun\ black hole in the 
nucleus, and provides an estimate of the upper limit on the black hole
mass.

\section{NGC 7469}

\begin{figure}
\centerline{\includegraphics[width=1.0\textwidth]{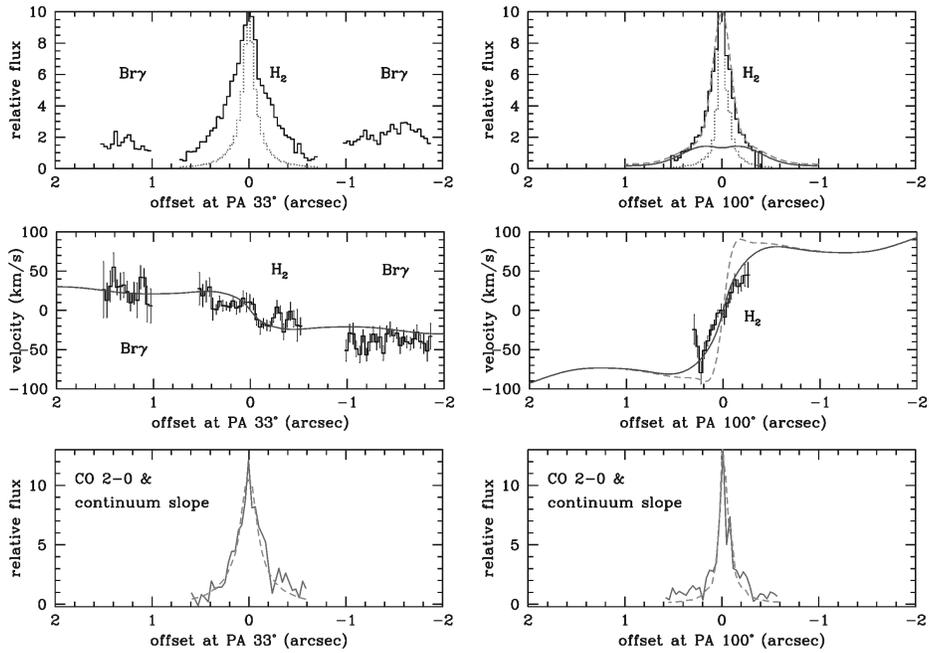}}
\caption{NGC\,7469. Upper: spatial profiles of the H$_2$ 1-0\,S(1) and
off-nuclear Br$\gamma$ lines at 2 position angles.
Continuum profiles are overdrawn (dotted lines).
At PA $100^\circ$ are also drawn the spatial profiles of the 2 models
shown in the centre panel.
Centre: velocity profiles of the emission lines.
Overdrawn are the best model (solid line), and at PA $100^\circ$ a
model which assumes that the mass surface density is traced by the
1-0\,S(1) line (pale dashed line).
Lower: spatial profiles of the stellar cluster, traced via CO\,2-0
bandhead absorption (solid line) and the continuum slope (dashed
line).}
\label{fig:ngc7469}
\end{figure}

The 85\,mas resolution spatial profiles of the 2.12\micron\ H$_2$
1-0\,S(1) line are 
plotted in the upper panels of Fig.~\ref{fig:ngc7469} for PAs
$33^\circ$ and $100^\circ$.
At $33^\circ$ we also show the circumnuclear 2.17\micron\ Br$\gamma$
line.
The velocity profiles of these emission lines are shown in
the centre panel.
Overplotted are the rotation curves for a model derived
from 0.7\arcsec\ radio CO\,2-1 data (solid dark curve, see
\cite{dav04a}).
This model consists of a broad disk component, a circumnuclear ring,
and a previously unknown nuclear ring at a radius of 0.2\arcsec;
it matches both the radio and near infrared data at their different
resolutions.
However, as shown in the right hand upper panel, the mass distribution
of the model is different to the profile of the 1-0\,S(1) line.
If instead, the 1-0\,S(1) line were to trace the mass
distribution (pale dashed curve) then the velocity gradient across the
inner 0.2\arcsec\ would be much steeper.
We conclude that the 1-0\,S(1) does not trace the mass distribution,
but that the emission probably originates in gas irradiated by X-rays
from the AGN.
Its morphology would then be dominated by the photon density and
dust distribution rather than the gas.

The lower panel in Fig.~\ref{fig:ngc7469} shows that we have resolved
the nuclear stellar cluster in NGC\,7469 using two different methods.
The solid line represents the spatial profile of the 2.29\micron\
CO\,2-0 bandhead absorption; the dashed line denotes the continuum
slope.
This latter method assumes that the continuum comprises stellar light
and hot dust emission, and decomposes the spectral slope at each point
accordingly.
Serendipitously, it demonstrates that the spatial resolution achieved
is in fact better than that measured directly from the full continuum.
We find that the stellar cluster is asymmetrical, with FWHM of
0.12\arcsec\ (30\,pc) and 0.22\arcsec\ (60\,pc) at the 2 PAs.
Using a combination of starburst models, the CO equivalent width, the
mass model, and the K-band flux density, we can show that in this region
the mass is dominated by stars rather than gas.
This result adds weight to our prior finding that the gas exists in a
nuclear ring around a compact star cluster, similar to the situation
seen in NGC\,1068 \cite{sch00,tha97}.

\section{Conclusions}

So far, this on-going study which uses
adaptive optics to probe the nuclear regions of AGN in the near
infrared on scales less than 0.1\arcsec, has shown that we can:
1) directly resolve nuclear star clusters, from scales of 30\,pc in
   NGC\,7469 to 300\,pc in Mkn\,231; 
2) understand the details of the dynamics in terms of thin
   disks, including bars, warps, rings, etc;
3) quantify the contributions to the mass
   from gas, stars, and the black hole itself on these small scales.


%


\begin{thebibliography}{}
\addcontentsline{toc}{section}{References}

\bibitem{dav04a} 
R. Davies, L. Tacconi, R. Genzel, 2004a, 
ApJ, in press

\bibitem{dav04b} 
R. Davies, L. Tacconi, R. Genzel, 2004b, 
to be submitted to ApJ

\bibitem{dow98}
D. Downes, P. Solomon, 1998,
ApJ, 507, 615

\bibitem{sch98}
E. Schinnerer, A. Eckart, L. Tacconi, 1998,
ApJ, 500, 147

\bibitem{sch00}
E. Schinnerer, A. Eckart, L. Tacconi, R. Genzel, 2000,
ApJ, 533, 850

\bibitem{sch03}
J. Scharw\"achter, A. Eckart, S. Pfalzner, J. Moultaka, 
C. Straubmeier, J. Staguhn, 2003, 
A\&A, 405, 959

\bibitem{tha97}
N. Thatte, A. Quirrenbach, R. Genzel, R. Maiolino, M. Tecza, 1997,
ApJ, 490, 238

\end{thebibliography}
\end{document}